\documentclass[pra,twocolumn,showpacs]{revtex4}
\usepackage{bm}
\usepackage{amstext}
\usepackage{rotating}
\usepackage{graphicx}
\usepackage{epsfig}
\usepackage{amssymb}
\usepackage{psfrag}

\begin{document}
\hyphenation{Fesh-bach}

\title{Binding Energies of $^{6}$Li $p$-wave Feshbach Molecules}

\author{J. Fuchs, C. Ticknor, P. Dyke, G. Veeravalli, E. Kuhnle, W. Rowlands, P. Hannaford, and C. J. Vale
}

\affiliation{ARC Centre of Excellence for Quantum Atom Optics,
Centre for Atom Optics and Ultrafast Spectroscopy,
Swinburne University of Technology, Melbourne, 3122, Australia\\}

\date{\today}

\begin{abstract}
We present measurements of the binding energies of $^6$Li $p$-wave Feshbach molecules formed in combinations of the $|F = 1/2, m_F = +1/2\rangle$ ($|1\rangle$) and $|F = 1/2, m_F = -1/2\rangle$ ($|2\rangle$) states.  The binding energies scale linearly with magnetic field detuning for all three resonances.  The relative molecular magnetic moments are found to be $113 \pm 7\,\mu$K/G, $111 \pm 6\,\mu$K/G and $118 \pm 8\,\mu$K/G for the $|1\rangle-|1\rangle$, $|1\rangle-|2\rangle$ and $|2\rangle-|2\rangle$ resonances, respectively, in good agreement with theoretical predictions.  Closed channel amplitudes and the size of the $p$-wave molecules are obtained theoretically from full closed-coupled calculations.

\end{abstract}


\maketitle

\section{Introduction}

Ultra-cold dilute Fermi gases provide an ideal system in which to probe aspects of pairing and superfluidity.  Rapid progress has been made through the use of magnetic field Feshbach resonances which can dramatically alter the two-body interactions.  These scattering resonances occur when the energy of two colliding atoms is Zeeman tuned to coincide with a bound molecular state.  Feshbach resonances play a key role in studies of superfluid Fermi gases and have led to the experimental realisation of the crossover from a Bose-Einstein Condensate (BEC) of molecules to a Bardeen-Cooper-Schrieffer (BCS) superfluid of long-range Cooper pairs~\cite{jochim03,greiner03,bourdel04,zwierlein03,regal04,partridge05}.

Of interest here is the extension of previous work on $s$-wave pairing to pairs with nonzero angular momentum.  Such gases display a BEC to BCS superfluid crossover and possess complex phase diagrams with phase transitions between different projections of the angular momentum \cite{botelho05,cheng05,gurarie05,iskin06,levinsen07}.  Condensates may also provide a link with other paired systems such as $d$-wave high-$\mathrm{T_C}$ cuprate superconductors~\cite{tsuei00} and liquid $^3$He~\cite{lee97}. Higher-order partial wave Feshbach resonances are intrinsically narrow in magnetic field because the interacting atoms must tunnel through the centrifugal barrier in order to interact.  Resonances which are narrow with respect to the Fermi energy can be accurately described in the low collision energy limit by a simple two-channel model~\cite{gurarie07}.

Initial experiments on $p$-wave Feshbach resonances in ultra-cold Fermi gases have focussed on $^6$Li and $^{40}$K.  In $^6$Li, three $p$-wave resonances have been identified in all combinations of atoms in the $|F = 1/2, m_F = +1/2\rangle$ ($|1\rangle$) and $|F = 1/2, m_F = -1/2\rangle$ ($|2\rangle$) states.  Atom loss associated with these resonances has been observed at fields of 159\,G, 185\,G and 215\,G due to the  $|1\rangle-|1\rangle$,  $|1\rangle-|2\rangle$ and $|2\rangle-|2\rangle$ resonances, respectively~\cite{zhang04, schunck05}.  Inelastic and elastic collision rates for the $|1\rangle-|2\rangle$ and $|2\rangle-|2\rangle$ resonances were calculated in \cite{chevy05}.  Evidence of molecule formation via adiabatically sweeping the magnetic field across the $|1\rangle-|2\rangle$ resonance was seen in \cite{zhang04}, but no long lived trapped molecules were detected.  Enhanced three-body loss has also been reported on the 159\,G $p$-wave resonance through interactions with a second species ($^{87}$Rb) \cite{deh08}.  Somewhat more progress has been made using $^{40}$K including measurements of the field dependent elastic scattering cross-section~\cite{regal03b} and the observation of the doublet corresponding to the different projections of the angular momentum \cite{ticknor04,gunter05}, where the weak dipole-dipole interaction lifts the degeneracy of the $m_l=\pm1$ and $m_l=0$ projections.  In 2007, Gaebler {\it{et al.}} succeeded in creating $p$-wave molecules from a gas of spin polarised $^{40}$K~\cite{gaebler07} using both magneto-association and three-body recombination and measured the binding energies and lifetimes in the bound and quasi-bound regimes.  Unfortunately, these molecules experience rapid decay to lower lying atomic spin states through dipolar relaxation and their lifetime was limited to less than 10\,ms.  This presents a major impediment to creating a $^{40}$K $p$-wave superfluid.  In $^6$Li, however, the $|1\rangle-|1\rangle$ $p$-wave resonance involves two atoms colliding in their lowest spin state, hence molecules produced on this resonance would not be susceptible to dipolar relaxation.  These molecules therefore have the potential to be much longer lived and may prove to be a viable avenue to studies of $p$-wave superfluidity with ultracold atomic gases.  Dipolar relaxation is also suppressed for the $m_l = -1$ projections of the $|1\rangle-|2\rangle$ and $|2\rangle-|2\rangle$ resonances due to angular momentum conservation.  However, these states are degenerate with other unstable $m_l$ projections, unlike the $|1\rangle-|1\rangle$ resonance which is stable for all projections.

In this paper, we present measurements of the binding energies of lithium $p$-wave Feshbach molecules formed on the three resonances in the two lowest hyperfine states.  A sinusoidally modulated magnetic field near a Feshbach resonance converts free atoms into bound or quasi-bound molecules.  Quasi-bound molecules are unique to atom pairs with nonzero angular momentum and possess positive energy, only being temporarily bound by the centrifugal barrier.  The rate of conversion depends on the resonant properties of the scattering states which we compare with theoretical predictions.  We also investigate the closed channel amplitude and size of $^6$Li $p$-wave molecules.

\section{Experiment}

Our experimental setup has been described previously \cite{fuchs07}; however, to increase the number of atoms that are cooled to degeneracy, atoms from the magneto-optical trap are loaded into a crossed optical dipole trap formed by a 100\,W fibre laser.  The two arms of the crossed dipole trap intersect at an angle of 14 degrees in a co-propagating geometry.  At full power, with 1/e$^2$ beam radii of approximately 40\,$\mu$m in both arms, a $\sim3$\,mK deep trap with oscillation frequencies of $\sim 11$\,kHz radially and $\sim 1.2$\,kHz axially results.  Approximately $10^6$ $^6$Li atoms are loaded into the crossed dipole trap in a near 50/50 spin mixture of the two lowest hyperfine states ($|1\rangle$ and $|2\rangle$).  Evaporative cooling is achieved by reducing the trap depth near the broad Feshbach resonance centred at 834\,G over $\sim 5$\,s.  In this way we can produce near pure Bose-Einstein condensates of approximately 60,000 $^6$Li$_2$ $s$-wave Feshbach molecules.

To probe narrow $^6$Li $p$-wave resonances high magnetic field stability and low field noise are required.  To achieve this we make use of two magnetic field coils.  A primary pair of coils, capable of providing fields up to 1.5\,kG, produces magnetic fields approximately 1\,G below the corresponding Feshbach resonance and a secondary pair of coils produces magnetic fields of up to a few Gauss and is used for fine tuning and fast switching.  The currents in both coils are actively stabilised by means of a feedback driven programmable power supply and MOSFET-switch, respectively.  At magnetic fields close to the $p$-wave resonances we achieve field a stability at the level of a few mG shot to shot and better than 20\,mG over the course of a few days.

\begin{figure}[t]
\begin{center}
\epsfig{file=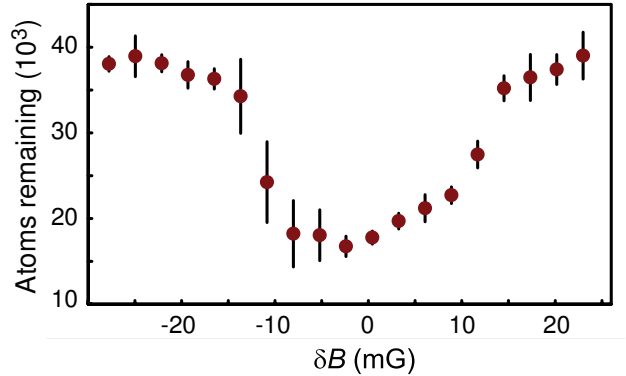,width=84mm}
\caption{Atom loss at the $|1\rangle$-$|1\rangle$ $^6$Li $p$-wave Feshbach resonance versus magnetic field detuning $\delta B = B_\mathrm{test} - B_0$ ($B_0 = 159$\,G).  The number of atoms remaining after 60\,ms hold time was measured via absorption imaging.  Each point is the average of six images. }
\label{fig:loss11}
\end{center}
\end{figure}

To characterise our system we have taken an atom loss measurement at the $|1\rangle-|1\rangle$ $p$-wave Feshbach resonance located at 159\,G.  A mixture of atoms in states $|1\rangle$ and $|2\rangle$ was evaporatively cooled at a magnetic field above the broad $s$-wave resonance to an energy of 150\,nK ($T \approx 0.2 T_F$).  Next the magnetic field was shifted well above resonance and atoms in state $|2\rangle$ were blasted away with a 40\,$\mu$s pulse of resonant laser light, which produced negligible heating of the remaining state $|1\rangle$ atoms.  The magnetic field was then switched to a value just above the $|1\rangle-|1\rangle$ $p$-wave resonance and the laser power lowered adiabatically in 20\,ms to yield a cloud energy of 100\,nK.  The magnetic field was subsequently ramped to a test value, $B_\mathrm{test}$, over $40$\,ms and then held there for $60$\,ms to map out the atom loss across the resonance.  Fig. \ref{fig:loss11} shows the atom number remaining after the hold time as a function of magnetic field detuning, $\delta B = B_\mathrm{test} - B_0$, where $B_0$ is 159.14\,G \cite{schunck05}.  Each data point is the average of six measurements.  The full width at half maximum of $\sim$ 25\,mG, is primarily limited by our field stability.  The splitting of the the $m_l = \pm1$ and $m_l=0$ projections is predicted to be 10\,mG for this resonance (compared to 500\,mG for $^{40}$K \cite{gaebler07}) which we cannot resolve in our current setup.  The asymmetry of the loss feature may be due to thermal or threshold effects and also the presence of the doublet.  As the $m_l = \pm1$ projections are themselves degenerate, there are twice as many possible scattering states for $|m_l| = 1$ than $m_l = 0$, and the latter occurs 10\,mG higher in field.

\section{Binding energies}

We have measured the binding energies, $E_B$, and obtained the magnetic moments, $\mu_m$, of lithium $p$-wave Feshbach molecules using magneto-association spectroscopy \cite{thompson05,gaebler07}.  By sinusoidally modulating the magnetic field close to a Feshbach resonance free atoms can be associated to bound or quasi-bound Feshbach molecules which are not seen in our absorption images.  The binding energies of $p$-wave molecules produced near the $|1\rangle-|1\rangle$ and $|2\rangle-|2\rangle$ Feshbach resonances were measured using spin-polarised gases of atoms at a temperature of $\sim$ 400\,nK in the appropriate state.  Binding energies of the $|1\rangle-|2\rangle$ molecules were measured at a temperature of $\sim$ 1\,$\mu$K to avoid producing $s$-wave molecules during the magnetic field ramp down to the $p$-wave field.  Measurements for the $|1\rangle-|1\rangle$ and $|2\rangle-|2\rangle$ molecules were performed in traps with final oscillation frequencies of $\sim 550$\,Hz radially and $\sim60$\,Hz axially, while for $|1\rangle-|2\rangle$ molecules a $\sim 870$\,Hz by $\sim95$\,Hz trap was used.

\begin{figure}[t]
\begin{center}
\epsfig{file=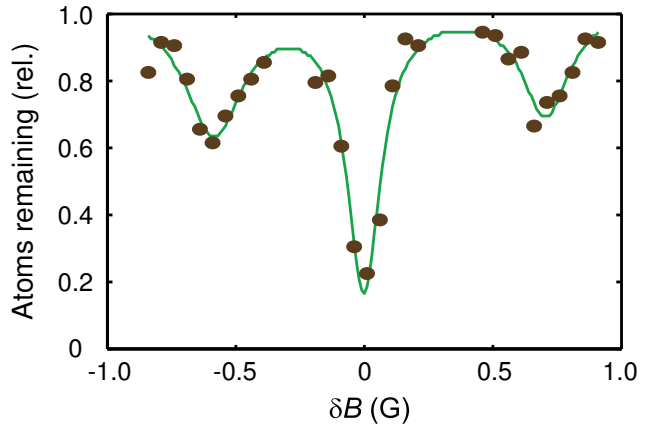,width=84mm}
\caption{Magneto-association spectrum for the $|2\rangle-|2\rangle$ $p$-wave Feshbach resonance in $^6$Li at $\nu_{\mathrm{mod}} = 650$\,kHz.  The central loss feature is due to three-body recombination on resonance while the loss features to the left and right are due to resonant magneto-association of bound and quasi-bound Feshbach molecules, respectively.}
\label{fig:rfscan}
\end{center}
\end{figure}

Atom-pair association occurs when the modulation frequency, $\nu_{\mathrm{mod}}$, corresponds to the energy difference between the free atom and bound or quasi-bound molecular states.  When the resonance condition is fulfilled, significant atom loss can be observed.  We apply the modulation by means of a 2.5\,cm diameter coil placed approximately 2\,cm below the atomic gas.  The oscillating magnetic field it produces is oriented along the same direction as the primary magnetic field and is held on for times, $t_{\mathrm{mod}}$, varying from  200\,ms to 2\,s.  At larger binding energies, longer modulation times were required to compensate for the reduced association rates.  The amplitude of the magnetic field modulation was kept fixed at 180\,mG for all experiments.  A typical scan is shown in Fig. \ref{fig:rfscan} where a spin-polarised gas of atoms in state $|2\rangle$ is probed close to the 215\,G Feshbach resonance.  The magnetic field was varied while the modulation frequency was fixed, in this instance at $\nu_{\mathrm{mod}} = $ 650\,kHz.  The principal loss feature in the centre of the scan is due to inelastic losses at the Feshbach resonance and its position coincides with the magnetic field where the free colliding atoms are degenerate with the molecular state. The two loss features on either side of this are due to magneto-association.  Atom loss on the low magnetic field side of the resonance is due to resonant conversion of atoms into {\it bound} Feshbach molecules.  The loss feature on the high magnetic field side is unique to Feshbach resonances involving molecules with nonzero angular momentum and is due to the production of {\it quasi-bound} molecules.  Quasi-bound pairs have energy above the free atom continuum and can tunnel through the centrifugal barrier, limiting their lifetimes to a few ms.  This leads to a broadening of the loss feature which was observed in $^{40}$K~\cite{gaebler07} and from which the lifetime of the quasi-bound molecules could be inferred.  In our experiments, however, we could not resolve any significant broadening of the line shape.  Furthermore, at the collision energies used in these experiments, we did not observe any noticeable temperature dependence of the position of the loss feature.  The $|m_l| = 1$ - $m_l = 0$ doublet was again unresolved in these spectra.  The solid line in Fig. \ref{fig:rfscan} is a fit of three Lorentzians from which we determine the central position of each loss feature.

\begin{figure}[t]
\begin{center}
\epsfig{file=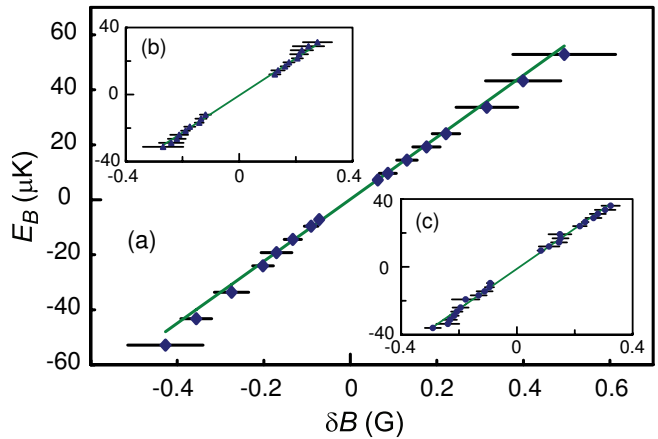,width=1\linewidth}
\caption{Binding energies of $p$-wave Feshbach molecules formed near the (a) $|1\rangle-|1\rangle$, (b) $|1\rangle-|2\rangle$ and (c) $|2\rangle-|2\rangle$ resonances.  All vertical (horizontal) axes are in units of $\mu$K (G).   Linear fits to these data yield gradients of (a) $113 \pm 7\,\mu$K/G, (b) $111 \pm 6\,\mu$K/G and (c) $118 \pm 8\,\mu$K/G.}
\label{fig:binding}
\end{center}
\end{figure}

By repeating these measurements using different $\nu_{\mathrm{mod}}$, it is possible to build up a picture of the binding energy as a function of magnetic field (ie. the magnetic moment of the molecules).  We have taken scans with up to eleven different modulation frequencies for each of the three $^6$Li $p$-wave Feshbach resonances and the fields corresponding to the bound and quasi-bound loss features are plotted in Fig. \ref{fig:binding}.  The main panel, (a), shows the binding energies for the $|1\rangle-|1\rangle$ resonance and the left, (b), and right, (c), insets are for the $|1\rangle-|2\rangle$ and $|2\rangle-|2\rangle$ resonances, respectively.  The binding energies vary linearly with magnetic field detuning and our measured gradients, listed in table 1, are in good agreement with our theoretical predictions for this resonance and previous work \cite{zhang04}.

\begin{center}
\begin{table}[h!]
%\vspace{0.5cm}

\begin{tabular}{|c|c|c|c|c|c|}
\hline
States & $B$\,(G) & $\Delta B$\,(mG) & $\mu_m^{exp}$ & $\: \mu_m^{th} \:$ & $\: \mu_m^{[15]} \:$ \\
\hline
$|1\rangle-|1\rangle$&159&10&$113\pm7$&113&-\\
$|1\rangle-|2\rangle$&185&4&$111\pm6$&116&117\\
$|2\rangle-|2\rangle$&215&12&$118\pm8$&111&111\\
\hline
\end{tabular}
\caption{Summary of the results of our binding energy measurements and calculations. For each of the three $p$-wave Feshbach resonances we give the approximate absolute magnetic field, $B$, the calculated splitting of the $|m_l| = 1$ - $m_l = 0$ doublet, $\Delta B$, our measured and calculated relative magnetic moments, $\mu_m^{exp,th}$, and the magnetic moment calculated in ref. \cite{zhang04}, $\mu_m^{[15]}$.  All magnetic moments are expressed in $\mu$K/G.}
\end{table}
\end{center}

Our theoretical binding energy slopes, $\mu_m^{th}$, are found from a full closed-coupled calculation \cite{weiner99} and are the average of the slopes on the bound and quasi-bound sides of the resonance.  On the bound side, these are approximately 0.7\,$\%$ steeper than on the quasi-bound side.  We are unable to resolve this difference within our experimental uncertainties.

We note that the slopes for the $^6$Li $p$-wave resonances are approximately twelve times steeper than measured for $^{40}$K $p$-wave molecules \cite{gaebler07}, making the requirements for magnetic field stability even more stringent.  The slope of the binding energy is due to the fact that, at these moderately high fields, the open channels are mostly triplet in character (having a magnetic moment near $2\mu_B$, where $\mu_B$ is the Bohr magneton), and the closed channel is predominantly spin singlet (with magnetic moment of zero).  The molecular state is closed-channel dominated (section \ref{sec:props}) and hence the relative magnetic moment of the molecules is close to $2\mu_B$.

\begin{figure}[t]
\begin{center}
\epsfig{file=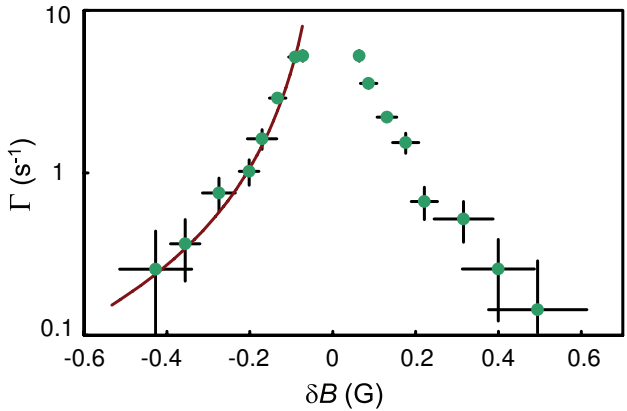,width=84mm}
\caption{Measured magneto-association rate, $\Gamma$, versus magnetic field detuning (points).  The solid line shows a scaled Fermi golden rule calculation for the transition rate from our calculated atomic and bound molecular states.  The scaling factor was chosen to fit the data and accounts for the density and temperature dependence of the conversion rate.}
\label{fig:lossrate}
\end{center}
\end{figure}

As the magnetic field detuning increases, the modulation time, $t_{\mathrm{mod}}$, required to see appreciable atom loss due to resonant magneto-association increases.  In the limit of low molecular conversion, the number of atoms remaining decays exponentially with time, $N_a(t) = N_{a0} e^{- \Gamma t}$, where $N_{a0}$ is the initial atom number and $\Gamma$ is the two-body association rate which varies with magnetic field.  We do not observe any Rabi oscillations in the atom number so fitting this exponential decay, for known $N_a(t)$ and $N_{a0}$, to our magneto-association spectra yields $\Gamma$ at the different magnetic fields.  We have done this for the data taken on the $|1\rangle-|1\rangle$ resonance and the results are plotted in Fig. \ref{fig:lossrate}.

Also shown in this figure is a scaled Fermi Golden Rule (FGR) calculation of the transition rate, $\Gamma_{th} \propto |\langle\psi_{mol}|\mu_z|\psi_{at}\rangle|^2$ where $\psi_{at}$ is the multi-channel wavefunction of the free atoms, $\psi_{mol}$ is the  multi-channel molecular wavefunction and $\mu_z$ is the magnetic dipole operator.  The real time dependent conversion rate depends on the density and temperature of the atomic cloud \cite{hanna07}.  As these were the same for each run of our experiments on a given resonance, these effects can be accounted for by a simple scaling factor which is the same for all runs.  The shape of the theoretical curve matches the experimental association rate extremely well over nearly two orders of magnitude, indicating that the states used in the FGR calculation are indeed a good representation of the true states.  The calculation was only performed on the bound side of the resonance as the localised molecular wavefunction allows the FGR integral to converge.

The molecular properties are roughly constant across the resonance (see below) and the FGR shows the change in the properties of the continuum scattering state near resonance.  We note that the experimental data is highly symmetric about $\delta B = 0$.  No evidence of increased conversion on the quasi-bound side of resonance is seen, even though the thermally averaged elastic scattering cross-section is known to be larger for $\delta B > 0$ \cite{regal03b,ticknor04}.

\section{Properties of the $^6$Li $p$-wave Feshbach molecules} \label{sec:props}
\begin{figure}
\centerline{\epsfig{file=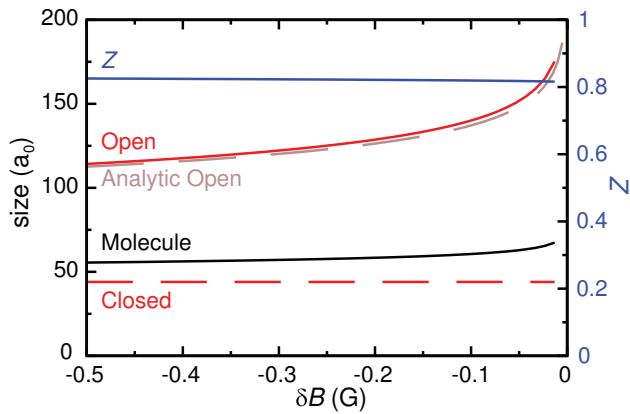,width=84mm}}
\caption{(Color Online)
Properties of $p$-wave Feshbach molecules as a function of
magnetic field detuning.  The size of the molecule (black), open (red)
and closed (dashed red) channel components are shown.  The closed
channel amplitude, Z (blue, right vertical axis), is shown as a function of detuning.}\label{zsize}
\end{figure}

We have also theoretically studied the properties of $p$-wave Feshbach molecules, and highlight some important distinctions from $s$-wave Feshbach molecules.  Most significantly, the $p$-wave molecular properties are essentially constant across the resonance, in strong contrast to $s$-wave molecules \cite{frmol}.  To illustrate this, first consider the closed channel amplitude, $Z$ \cite{frmol,stoof}.  This entails writing the wavefunction as
\begin{eqnarray}
|\psi_{mol}\rangle=\sqrt{Z}|\psi_c\rangle +\sqrt{1-Z}|\psi_o\rangle, \label{ocz}
\end{eqnarray}
where $\psi_o$ is the open channel component, and $\psi_c$ represents all closed channel components.  Physically, the closed channels correspond to channels with different (separate) atomic hyperfine quantum numbers which are energetically forbidden at long range, but couple to the open channel at short range via spin-exchange.  In Fig. \ref{zsize} we have plotted $Z$ (solid blue) for $^6$Li $p$-wave molecules as a function of detuning (right vertical axis) on the bound side of the resonance.  $Z$ was obtained directly from the full closed-coupled calculation and is roughly 0.82 for all detunings shown.  This property is essentially constant across the resonance until the detuning is extremely small ($<5$\,mG), beyond our current experimental resolution.

With the magnetic moment of the molecule, we can estimate the population of the closed channel experimentally from $Z = {1\over\Delta\mu}{dE_b\over dB} = {\mu_m\over\Delta\mu}$ where $\Delta\mu$ is the relative magnetic moment of the open and closed channels \cite{frmol}.  Using only data from the molecular side of the resonance ($\delta B < 0$) yields $\mu_m = 115 \pm 9$\,$\mu$K/G.  From our closed coupled calculations we find $\Delta \mu$ = 2.12$\mu_b =142\,\mu$K/G giving an experimental value of $Z = 0.81 \pm 0.07$, in good agreement with the theory (0.82).  $Z$ is important because it can be related to other molecular properties, such as the molecular size.

%+++++++++++++++++++++++++++++++++++++++++++++++++++++++++++++++++++++

In Fig. \ref{zsize} we have also plotted the size of the $p$-wave molecules (black) as a function of detuning.  In addition we have separately plotted the size of a dominant closed channel (red dashed), the open channel (red), and an analytic open channel model (dashed brown, see below).  We can understand this behavior by expressing the size of the molecule in terms of open and closed channels:
\begin{eqnarray} \label{eq:r}
r_{mol} = Zr_c + (1-Z)r_o
\end{eqnarray}
where $r_i=\langle \psi_i|r|\psi_i\rangle$, and $i=\{mol,o,c\}$. This shows that the size of the molecule very closely follows $Z$.  In addition to this, both $r_o$ and $r_c$ are much smaller than those in the $s$-wave molecules.  This fact arises due to the centrifugal barrier confining the wavefunction to short range.  To see this more clearly, we compare the asymptotic forms of both the $s$-wave and $p$-wave radial wavefunctions:
\begin{eqnarray}
&&\psi_{o}^{s}(r) \propto {e^{-r/a}\over r} \label{psims} \\
&&\psi_{o}^{p}(r) \propto {e^{-kr}\over r}\left(1+{1\over kr}\right) \label{psimp}
\end{eqnarray}
where $a$ is the $s$-wave scattering length.  Near an $s$-wave resonance the molecules are open channel dominated because $Z \rightarrow 0$ as $\delta B \rightarrow 0$ \cite{frmol}.  This means $r_{mol}^s \sim r_o$ (Eq. \ref{eq:r}).  From Eq. \ref{psims} one finds $r_o$ is remarkably large for $s$-wave molecules: $r_{mol}^s=a/2$ and $a= a_{bg}(1-\Delta/\delta B)$ where $a_{bg}$ is the off resonant scattering length and $\Delta$ is the width of the Feshbach resonance.  In contrast, $p$-wave molecules carry much more amplitude at short range due to the extra radial dependence in Eq. \ref{psimp}.  To illustrate this, one can use Eq. \ref{psimp} to make an analytic approximation to the size of the open channel using $\hbar k= \sqrt{2 m E_B} = \sqrt{M \mu_m |\delta B|}$, where $m$ is the reduced mass, equal to half the atomic mass, $M$.  The analytic model diverges as $r \rightarrow 0$; so we choose a cut off radius of $35a_0$, such that the model and full calculation have similar $r_o$ values at large detuning.  $r_o$ from the analytic model is shown as a dashed brown line in Fig. \ref{zsize}.  Note that even as the detuning becomes small ($\delta B \approx 5$\,mG) the size of the open channels remains quite small, $r_o < 200a_0$.  Additionally, $p$-wave moleucules are closed-channel dominated ($Z \sim$ 0.8), and $r_c \approx 40a_0$ is constant for all $\delta B$.  The small size of the open channel, combined with the closed-channel character of $p$-wave molecules, results in $r_{mol}^p < 70a_0$, which is much smaller than typical $s$-wave Feshbach molecules.

It is interesting to consider what implications the molecular size has for the realisation of a BEC-BCS crossover regime for nonzero orbital angular momentum pairing.  The crossover regime for $s$-wave pairs is comparatively smooth because the molecular size grows appreciably as the detuning approaches zero from below.  However, for $p$-wave pairs the crossover will be much more abrupt as the molecular size barely changes at the resonance.  Additionally, on the BCS side of a $p$-wave resonance, the pair wavefunction may have significant amplitude at short range because of the centrifugal barrier.  We also expect an increase in the rate of inelastic vibrational quenching collisions between molecules and free atoms which release large amounts of energy and lead to rapid loss.  Fermionic particle statistics greatly suppresses this process for weakly bound $s$-wave molecules comprised of two fermions \cite{petrov04}.  However, near a $p$-wave Feshbach resonance, fermions in the same spin state interact resonantly so this suppression mechanism, and the considerations leading to it, will not apply.  In fact, it has recently been shown that $p$-wave molecule-atom collisions are expected to be largely insensitive to the two-body $p$-wave elastic cross-section which can lead to short lifetimes of an atom-molecule mixture \cite{dincao08}.

\section{Conclusion}

In this paper we have presented measurements of the binding energies of the three $p$-wave Feshbach molecules formed by $^6$Li atoms in the two lowest hyperfine states.  We find that the binding energy scales approximately linearly with the magnetic field detuning.  Our data agrees well with theoretical calculations for the binding energy slopes, implying agreement for the closed channel amplitude and molecule size.  The techniques and results presented here may provide a foundation towards the production and observation of long-lived $p$-wave Feshbach molecules made up from fermionic atoms.  To date we have not observed any evidence of $^6$Li $|1\rangle-|1\rangle$ $p$-wave molecules remaining trapped after production via either rf association or three-body recombination.  In our experiments molecule production takes place over times of order tens of ms and absorption imaging takes place immediately after this.  After submission of this paper we became aware of related work in which $^6$Li $|1\rangle-|1\rangle$ $p$-wave molecules were produced and detected \cite{inada08} and the elastic and inelastic collision rates were measured.  In the presence of unpaired atoms, $p$-wave molecules were found to decay within a few ms via vibrational quenching collisions.  In our experiments, unpaired atoms were present which explains why we did not observe molecules trapped after a few tens of ms.  A significant increase in lifetime is possible after removing the free atoms \cite{inada08}.  Provided the rate of vibrational quenching can be controlled $^6$Li $p$-wave molecules produced on the $|1\rangle-|1\rangle$ resonance may still be a promising candidate for achieving superfluidity of atom pairs with nonzero angular momentum in dilute gases.

\acknowledgments
This project is supported by the Australian Research Council Centre of Excellence for
Quantum-Atom Optics and Swinburne University of Technology strategic initiative funding.

\bibliographystyle{amsplain}

\end{document}